\documentclass[twocolumn,aps,showpacs,superscriptaddress]{revtex4-1}
\usepackage{amsmath}
\usepackage{graphicx}
\RequirePackage[pagewise,mathlines]{lineno}

\begin{document}
\title{Coherent photo-produced J$/\psi$ and dielectron yields in isobaric collisions}%
\author{W. Zha}\email{first@ustc.edu.cn}\affiliation{University of Science and Technology of China, Hefei, China}
\author{L. Ruan}\affiliation{Brookhaven National Laboratory, New York, USA}
\author{Z. Tang}\email{zbtang@ustc.edu.cn}\affiliation{University of Science and Technology of China, Hefei, China}
\author{Z. Xu}\affiliation{Brookhaven National Laboratory, New York, USA}\affiliation{Shandong University, Jinan, China}
\author{S. Yang}\affiliation{Brookhaven National Laboratory, New York, USA}
\date{\today}%
\begin{abstract}
Recently, significant enhancements of J/$\psi$ and $e^{+} e^{-}$ pair production at very low transverse momenta were observed by the STAR and ALICE collaboration in peripheral hadronic A+A collisions. The anomaly excesses point to evidence of coherent photon-nucleus and photon-photon interactions in violent hadronic heavy-ion collisions, which were conventionally studied only in ultra-peripheral collisions. The isobaric collisions performed at RHIC provides a unique opportunity to test the existence of coherent photon products in hadronic heavy-ion collisions. The idea is that the possible production of coherent photon products is significantly different in different collision systems due to the variations in their charge and nuclear density distributions. In this letter, we focus on the peripheral collisions and provide theoretical predictions for coherent production of J/$\psi$ and dielectron in isobaric collisions. We show that the expected yields differ significantly to perform the experimental test.
\end{abstract}
\maketitle
\section{Introduction}
In relativistic heavy-ion collisions carried out at the Relativistic Heavy Ion Collider (RHIC) and the Large Hadron Collider (LHC), one aims at searching for a new form of matter --- the Quark-Gluon Plasma (QGP)~\cite{PBM_QGP} and studying its properties in laboratory. The production of J/$\psi$ and dileptons in heavy-ion collisions are key measurements to probe the formation of QGP. Due to the color screening of the quark-antiquark potential in the deconfined medium, the production of J/$\psi$ would be significantly suppressed, which was proposed as a direct signature of the QGP formation~\cite{MATSUI1986416}. After decades of experimental and theoretical efforts, it is recognized that other mechanisms, such as the recombination of deconfined charm quarks in the QGP and cold nuclear matter (CNM) effects, also modify the J/$\psi$ production significantly in heavy-ion collisions. Currently, the interplay of these effects can qualitatively explain the J/$\psi$ yield measured so far at SPS, RHIC and LHC~\cite{STAR_Jpsi_AuAu_BES}. Dileptons have been proposed as ``penetrating probes'' for the hot and  dense medium~\cite{SHURYAK198071}, because they are not subject to the violent strong interactions in the medium. Various dilepton measurements have been performed in heavy-ion collisions. A clear enhancement in low mass region ($M_{ll} < M_{\phi}$) has been observed, which is consistent with in-medium broadening of the $\rho$ mass spectrum~\cite{PhysRevLett.79.1229,KOHLER2014665}. While the excess presented in the intermediate mass region ($M_{\phi} < M_{ll} < M_{J/\psi}$) is believed to be originated from the QGP thermal radiation~\cite{PhysRevC.63.054907}.

J/$\psi$ and dilepton can also be generated by the intense electromagnetic fields accompanied with the relativistic heavy ions~\cite{UPCreview}. The intense electromagnetic field can be viewed as a spectrum of equivalent photons by the equivalent photon approximation~\cite{KRAUSS1997503}. The quasi-real photon emitted by one nucleus could fluctuate into $c\bar{c}$ pair, scatters off the other nucleus, and emerge as a real J/$\psi$. The virtual photon from one nucleus can also interact with the photon from the other, resulting in the production of dileptons, which can be represented as $\gamma + \gamma \rightarrow l^{+} + l^{-} $. The coherent nature of these interactions gives the processes distinctive characteristics: the final products consist of a J/$\psi$ (or dilepton pair) with very low transverse momentum, two intact nuclei, and nothing else. Conventionally, these reactions are only visible and studied in Ultra-Peripheral Collisions (UPC), in which the impact parameter ($b$) is larger than twice the nuclear radius ($R_{A}$) to avoid any hadronic interactions.

Can the coherent photon products also exist in Hadronic Heavy-Ion Collisions (HHIC, $b < 2R_{A}$), where the violent strong interactions occur in the overlap region? The story starts with the measurements from ALICE: significant excesses of J/$\psi$ yield at very low $p_{T} (< 0.3$ GeV/c) have been observed in peripheral Pb+Pb collisions at $\sqrt{s_{\rm{NN}}} =$ 2.76 TeV~\cite{LOW_ALICE}, which can not be explained by the hadronic J$/\psi$ production with the known cold and hot medium effects. STAR made the same measurements in Au+Au collisions at $\sqrt{s_{\rm{NN}}} =$ 200 GeV~\cite{1742-6596-779-1-012039}, and also observed significant enhancement at very low $p_{T}$ in peripheral collisions. The anomaly excesses observed possess characteristics of coherent photoproduction and can be quantitative described by the theoretical calculations with coherent photon-nucleus production mechanism~\cite{PhysRevC.93.044912,PhysRevC.97.044910,SHI2018399}, which points to evidence of coherent photon-nucleus reactions in HHIC. If coherent photonuclear production is the underlying mechanism for the observed J/$\psi$ excess, coherent photon-photon production should also be there and contribute to the dilepton pair production in HHIC. Base on this train of thought, STAR measured the dielectron spectrum at very low $p_{T}$ in peripheral collisions, and indeed significant excesses were observed~\cite{PhysRevLett.121.132301}, which could be reasonably described by coherent photon-photon production mechanism~\cite{ZHA2018182,PhysRevC.97.054903}. The isobaric collision experiment, recently completed in the 2018 run at RHIC ($^{96}_{44}\rm{Ru} + ^{96}_{44}\rm{Ru}$ and $^{96}_{40}\rm{Zr} + ^{96}_{40}\rm{Zr}$), provides a unique opportunity to further address this issue. The idea is that: the production yield originated from coherent photon-nucleus interaction should be proportional to $Z^{2}$, while the production rate of coherent photon-photon is proportional to $Z^{4}$; according to above, the excesses of J$/\psi$ and dielectron would differ significantly between isobaric collisions and Au+Au collisions for centralities with the same hadronic background. In this letter, we report calculations for coherent production of J/$\psi$ and dielectron in isobaric collisions to provide theoretical baseline for further experimental test. The centrality dependence of the coherent products is presented and compared to that in Au+Au collisions. The difference in $t$ distributions of J/$\psi$ between isobaric collisions and Au+Au collision is also discussed in current framework.
\section{Theoretical formalism}
According to the equivalent photon approximation, the coherent photon-nucleus and photon-photon interactions in heavy-ion collisions can be factorized into a semiclassical and quantum part. The semiclassical part deals with the distribution of quasi-real photons induced by the colliding ions, while the quantum part handles the interactions of photon-Pomeron or photon-photon. The cross section for J/$\psi$ from coherent photon-nucleus interactions can be written as~\cite{UPC_JPSI_PRC,UPC_JPSI_PRL}:
  \begin{equation}
  \label{equation1}
  \sigma({A + A} \rightarrow {A + A} + \text{J}/\psi) = \int d\omega n(\omega)\sigma(\gamma A \rightarrow \text{J}/\psi A),
  \end{equation}
  where $\omega$ is the photon energy, $n(\omega)$ is the photon flux at energy $\omega$, and $\sigma(\gamma A \rightarrow \text{J}/\psi A)$ is the photonuclear interaction cross-section for J$/\psi$.
 Similarly, the dielectron production from coherent photon-photon reactions can be calculated via~\cite{Klein:2016yzr}:
     \begin{equation}
    \begin{aligned}
    &\sigma (A + A \rightarrow A + A + e^{+}e^{-})
    \\
    & =\int d\omega_{1}d\omega_{2} \frac{n(\omega_{1})}{\omega_{1}}\frac{n(\omega_{2})}{\omega_{2}}\sigma(\gamma \gamma \rightarrow e^{+}e^{-}),
         \label{equation2}
    \end{aligned}
    \end{equation}
    where $\omega_{1}$ and $\omega_{2}$ are the photon energies from the two colliding beams, and $\sigma(\gamma \gamma \rightarrow e^{+}e^{-})$ is the photon-photon reaction cross-section for dielectron.

\begin{figure*}[htbp]
\includegraphics[keepaspectratio,width=0.45\textwidth]{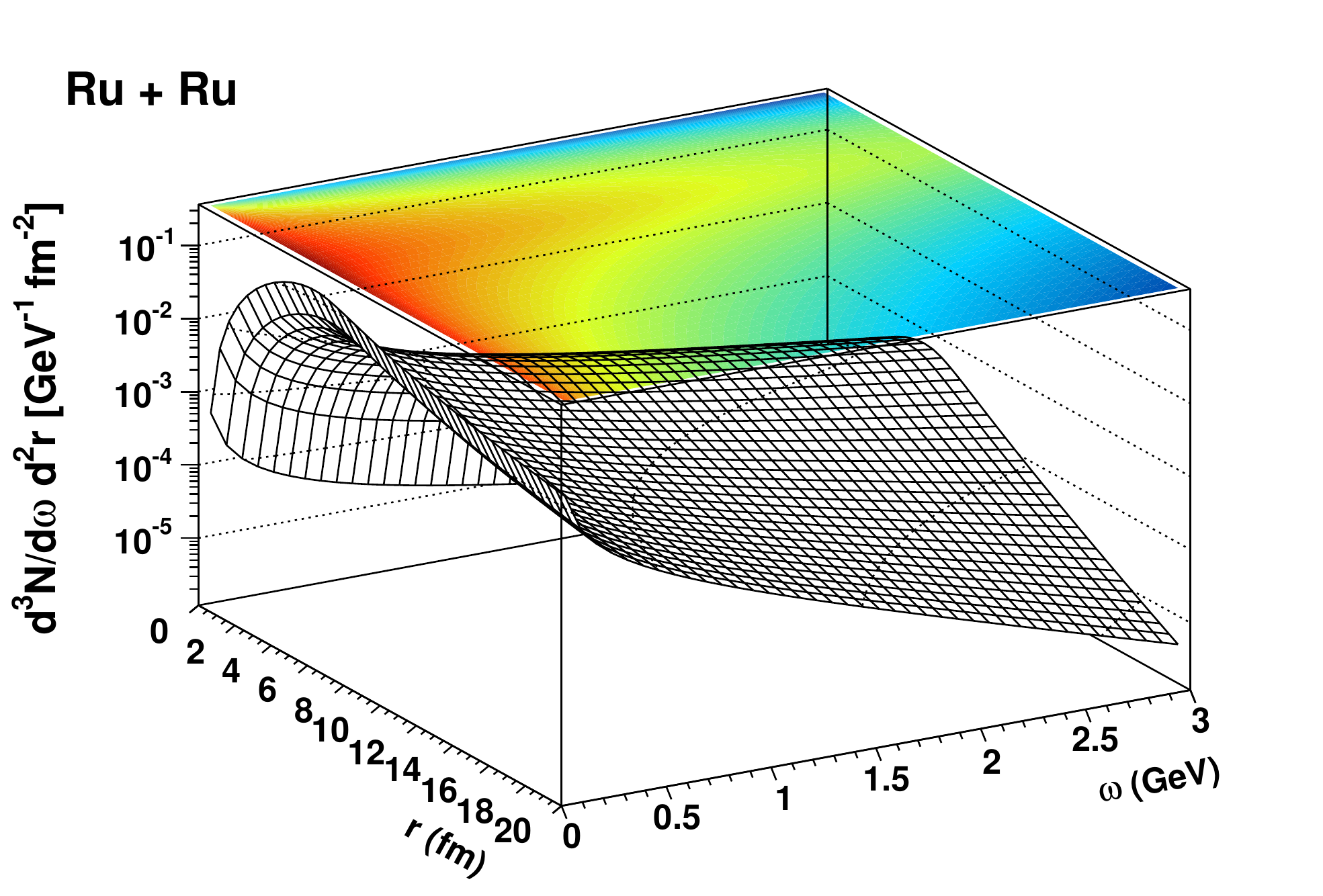}
\includegraphics[keepaspectratio,width=0.45\textwidth]{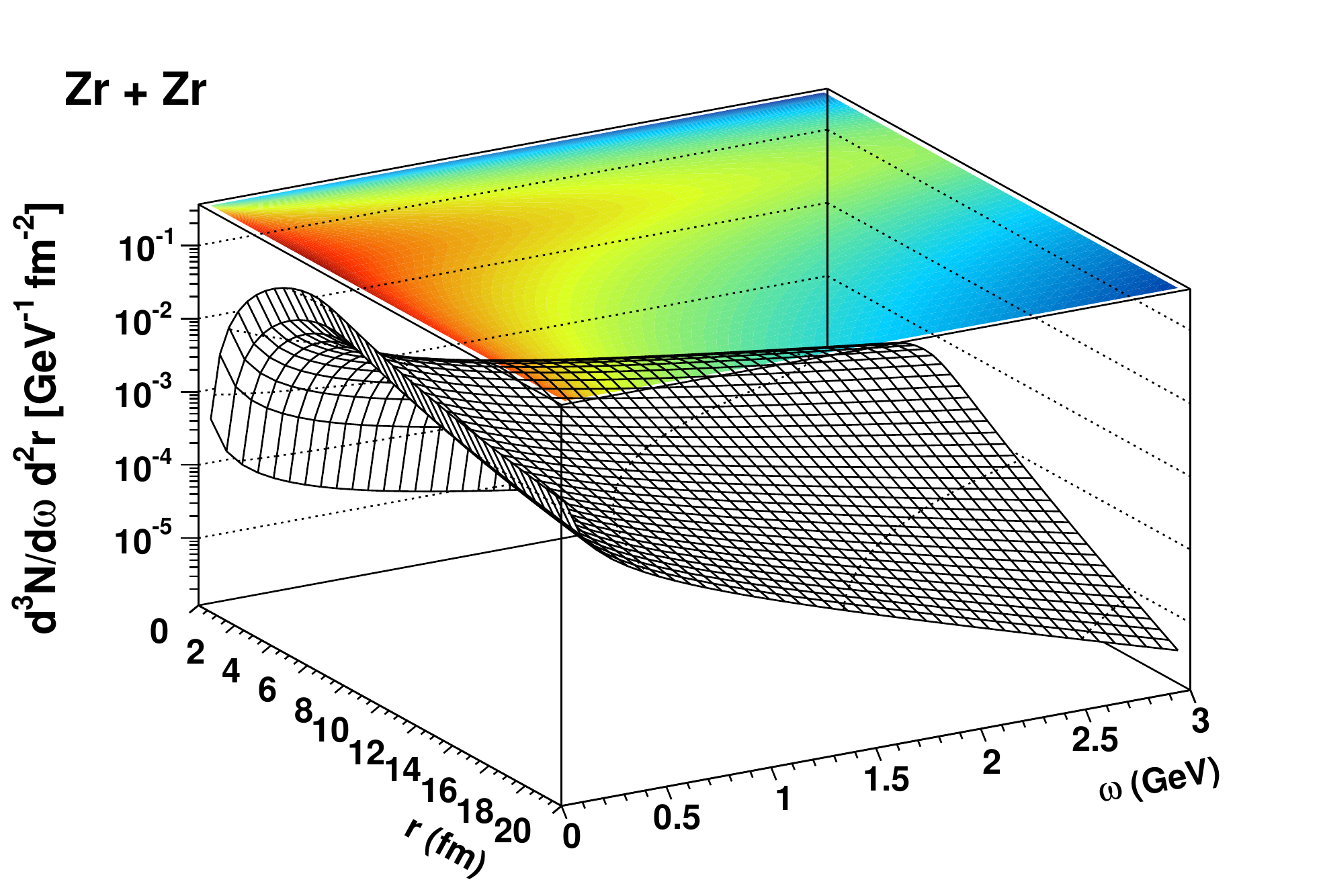}
\caption{Two-dimensional distributions of the photon flux in the distant $r$ and in the energy of photon $\omega$ for Ru + Ru (left panel) and Zr + Zr (right panel) collisions at $\sqrt{s_{\rm{NN}}} =$ 200 GeV.}
\label{figure1}
\end{figure*}

    The photon flux induced by the heavy ions can be modelled using the Weizs\"acker-Williams method~\cite{KRAUSS1997503}:
       \begin{equation}
  \label{equation3}
  \begin{aligned}
  & n(\omega,r) = \frac{4Z^{2}\alpha}{\omega} \bigg | \int \frac{d^{2}q_{\bot}}{(2\pi)^{2}}q_{\bot} \frac{F(q)}{q^{2}} e^{iq_{\bot} \cdot r} \bigg |^{2}
  \\
  & q = (q_{\bot},\frac{\omega}{\gamma})
  \end{aligned}
  \end{equation}
  where $n(\omega,r)$ is the flux of photons with energy $\omega$ at distant $r$ from the center of nucleus, $\alpha$ is the electromagnetic coupling constant,$\gamma$ is lorentz factor, and the form factor $F(q)$ is Fourier transform of the charge distribution in nucleus. In the calculations, we employ the Woods-Saxon form to model the nucleon distribution of nucleus in spherical coordinates:
   \begin{equation}
  \rho_{A}(r,\theta)=\frac{\rho^{0}}{1+\exp[(r-R_{\rm{WS}}-\beta_{2}R_{\rm{WS}}Y_{2}^{0}(\theta))/d]},
  \label{equation4}
  \end{equation}
  where $\rho_{0} = 0.16 \rm{\ fm}^{-3}$, $R_{WS}$ and $d$ are the ``radius'' and the surface diffuseness parameter, respectively, and $\beta_{2}$ is the deformity of the nucleus. The deformity parameter $\beta_{2}$ for Ru and Zr is ambiguous and important for the bulk correlated physics~\cite{PhysRevC.97.044901}, however, it is a minor effect in our calculations. For simplicity, the deformity parameter $\beta_{2}$ is ignored and set to 0. The ``radius'' $R_{WS}$ (Au: 6.38 fm, Ru: 5.02 fm, Zr: 5.02fm) and surface diffuseness parameter $d$ (Au: 0.535 fm, Ru: 0.46 fm, Zr: 0.46 fm) are based on fits to electron scattering data~\cite{0031-9112-29-7-028}. Fig.~\ref{figure1} shows the two-dimensional distributions of the photon flux induced in isobaric collisions at $\sqrt{s_{\rm{NN}}} =$ 200 GeV as a function of distant $r$ and energy $\omega$ for Ru + Ru (left panel) and Zr + Zr (right panel).

 \renewcommand{\floatpagefraction}{0.75}

The cross-section for $\gamma A \rightarrow \text{J}/\psi A$ reaction can be derived from a quantum Glauber approach coupled with the parameterized forward scattering cross section $\frac{d\sigma(\gamma p \rightarrow \text{J}/\psi p)}{dt}|_{t=0}$ as input~\cite{PhysRevC.93.044912,UPC_JPSI_PRC,PhysRevC.97.044910}:
\begin{equation}
    \label{equation3_1}
    \begin{split}
    &\sigma(\gamma A \rightarrow \text{J}/\psi A)=\frac{d\sigma(\gamma A \rightarrow \text{J}/\psi A)}{dt}\bigg|_{t=0} \times\\
     &\int|F_{P}(\vec{k}_{P})|^{2}d^{2}{\vec{k}_{P\bot}} \ \ \ \ \ \ \vec{k}_{P}=(\vec{k}_{P\bot},\frac{ \omega_{P}}{\gamma_{c}})\\
  & \omega_{P} = \frac{1}{2}M_{\text{J}/\psi} e^{\pm y} = \frac{M_{\text{J}/\psi}^{2}}{4\omega_{\gamma}}
    \end{split}
    \end{equation}
    \begin{equation}
    \label{equation3_2}
    \frac{d\sigma(\gamma A \rightarrow \text{J}/\psi A)}{dt}\bigg|_{t=0}=C^{2}\frac{\alpha \sigma_{tot}^{2}(\text{J}/\psi A)}{4f_{\text{J}/\psi}^{2}}
    \end{equation}
    \begin{equation}
    \label{equation3_3}
    \sigma_{tot}(\text{J}/\psi A)=2\int(1-\exp(-\frac{1}{2}\sigma_{tot}(\text{J}/\psi p)T_{A}(x_{\bot})))d^{2}x_{\bot}
    \end{equation}
    \begin{equation}
    \label{equation3_4}
    \sigma_{tot}^{2}(\text{J}/\psi p)=16\pi\frac{d\sigma(\text{J}/\psi p \rightarrow \text{J}/\psi p)}{dt}\bigg|_{t=0}
    \end{equation}
    \begin{equation}
    \label{equation3_5}
    \frac{d\sigma(\text{J}/\psi p \rightarrow \text{J}/\psi p)}{dt}\bigg|_{t=0}=\frac{f_{\text{J}/\psi}^{2}}{4\pi \alpha C^{2}}\frac{d\sigma(\gamma p \rightarrow \text{J}/\psi p)}{dt}\bigg|_{t=0}
    \end{equation}
where $T_{A}(x_{\bot})$ is the nuclear thickness function, $-t$ is the squared four momentum transfer, and $f_{\text{J}/\psi}$ is the J/$\psi$-photon coupling. Eq.~\ref{equation3_2} and ~\ref{equation3_5} are relations from vector meson dominance model~\cite{RevModPhys.50.261} and the correction factor $C$ is adopted to account for the non-diagonal coupling through higher mass vector mesons~\cite{HUFNER1998154}, as implemented in the generalized vector dominance model~\cite{PhysRevC.57.2648}. Eq.~\ref{equation3_4} is the optical theorem relation and parametrization for forward cross section $\frac{d\sigma(\gamma p \rightarrow \text{J}/\psi p)}{dt}|_{t=0}$ in Eq.~\ref{equation3_5} is obtained from~\cite{Klein:2016yzr}.

The elementary cross-section to produce a pair of positron-electron with electron mass $m$ and pair invariant mass $W$ can be determined by the Breit-Wheeler formula~\cite{PhysRevD.4.1532}
     \begin{equation}
  \label{equation5}
  \begin{aligned}
  & \sigma (\gamma \gamma \rightarrow l^{+}l^{-}) =
  \\
  &\frac{4\pi \alpha^{2}}{W^{2}} [(2+\frac{8m^{2}}{W^{2}} - \frac{16m^{4}}{W^{4}})\text{ln}(\frac{W+\sqrt{W^{2}-4m^{2}}}{2m})
  \\
  & -\sqrt{1-\frac{4m^{2}}{W^{2}}}(1+\frac{4m^{2}}{W^{2}})].
  \end{aligned}
  \end{equation}
The angular distribution of these positron-electron pairs can be given by
 \begin{equation}
  G(\theta) = 2 + 4(1-\frac{4m^{2}}{W^{2}})\frac{(1-\frac{4m^{2}}{W^{2}})\text{sin}^{2}(\theta)\text{cos}^{2}(\theta)+\frac{4m^{2}}{W^{2}}}{(1-(1-\frac{4m^{2}}{W^{2}})\text{cos}^{2}(\theta))^{2}},
  \label{equation6}
  \end{equation}
 where $\theta$ is the angle between the beam direction and one of the electrons in the positron-electron center of mass frame.

 The approaches employed in this calculation are very mature in UPC, which could quantitatively describe the experimental measurements~\cite{SHI2018399,PhysRevC.70.031902,DYNDAL2017281,Abbas2013,2017489}. However, the energetic strong interactions in HHIC could impose significant impact on the coherent production. The possible disruptive effect could be factorized into two distinct sub-process: photon emission and external disturbance in overlap region. The equivalent photon field is highly contracted into the transverse plane, and travels along with the colliding nuclei in the laboratory frame. Therefore the coherent photon-nucleus and photon-photon interactions occur at almost the same time when violent hadronic collisions happen. Due to the time retarded potential, the quasi-real photons are likely to be emitted before hadronic collision by about $\Delta t = \gamma R/c$, where $\gamma$ is Lorentz factor, and $R$ is the transverse distance from the colliding nuclei. Hence, the photon emission should be unaffected by hadronic collisions. In the overlap region of collisions, the photon products could be affected by the violent hadronic interactions, leading to the loss of coherent action. For the coherent photon-photon interactions, because the final product is electron-positron pair, which is not subject to the strong interactions, the disruption effect from overlap region should be small enough to be neglected in the calculations. However, the J/$\psi$'s from coherent photon-nucleus interactions are sensitive to the hadronic interactions, thus the production in overlap region is prohibited in our approach. Furthermore, as described in Ref.~\cite{PhysRevC.97.044910}, the interference effect is included in the calculations of coherent photon-nucleus process.
 \section{Results}
  \renewcommand{\floatpagefraction}{0.75}
\begin{figure}[htbp]
\includegraphics[keepaspectratio,width=0.45\textwidth]{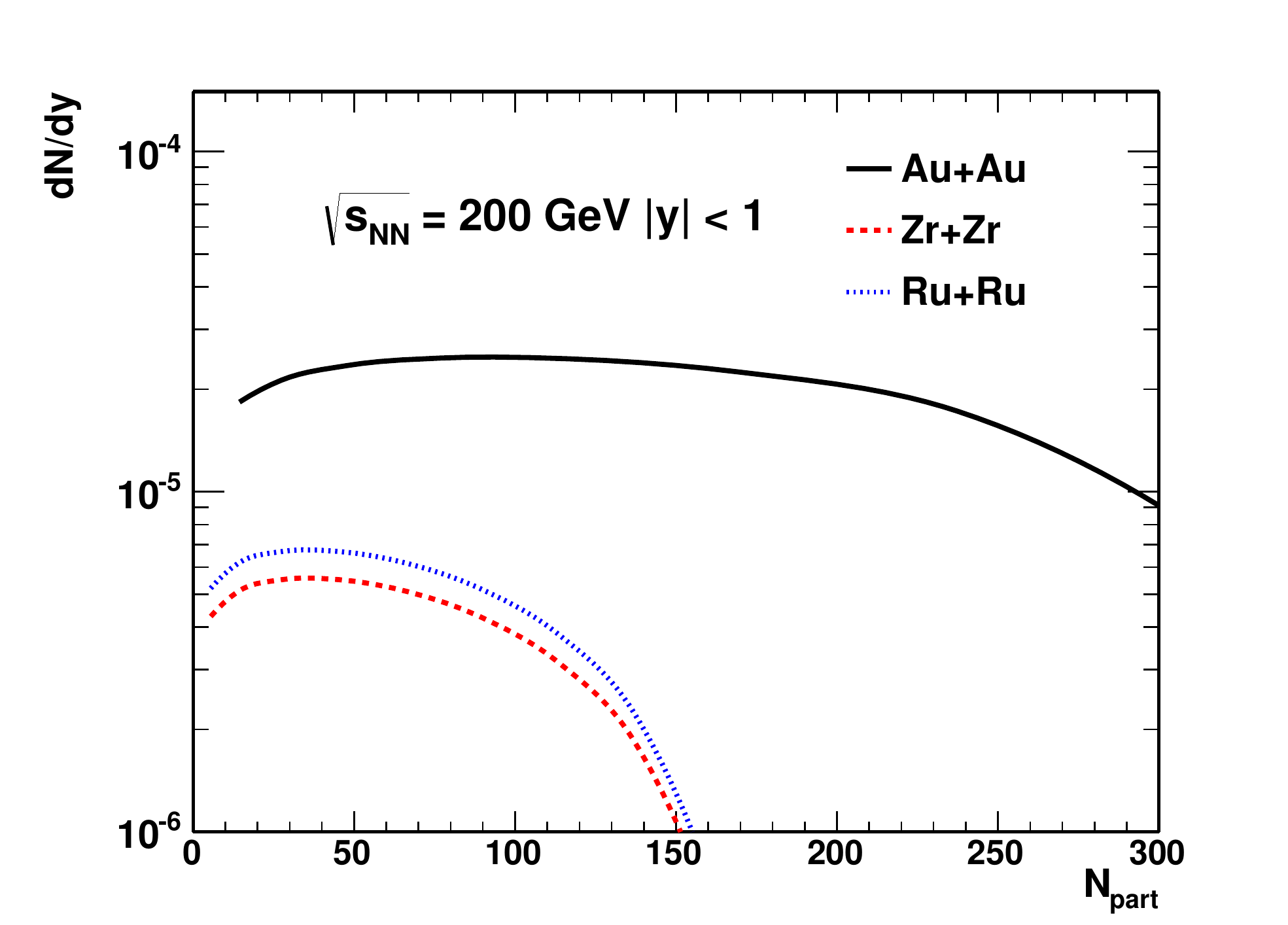}
\caption{Yields of coherent J/$\psi$ production as a function of $N_{\rm{part}}$ at $\sqrt{s_{\rm{NN}}} =$ 200 GeV in Au+Au (solid line), Ru+Ru (dotted line), and Zr+Zr (dashed line) collisions.}
\label{figure2}
\end{figure}

  \renewcommand{\floatpagefraction}{0.75}
\begin{figure}[htbp]
\includegraphics[keepaspectratio,width=0.45\textwidth]{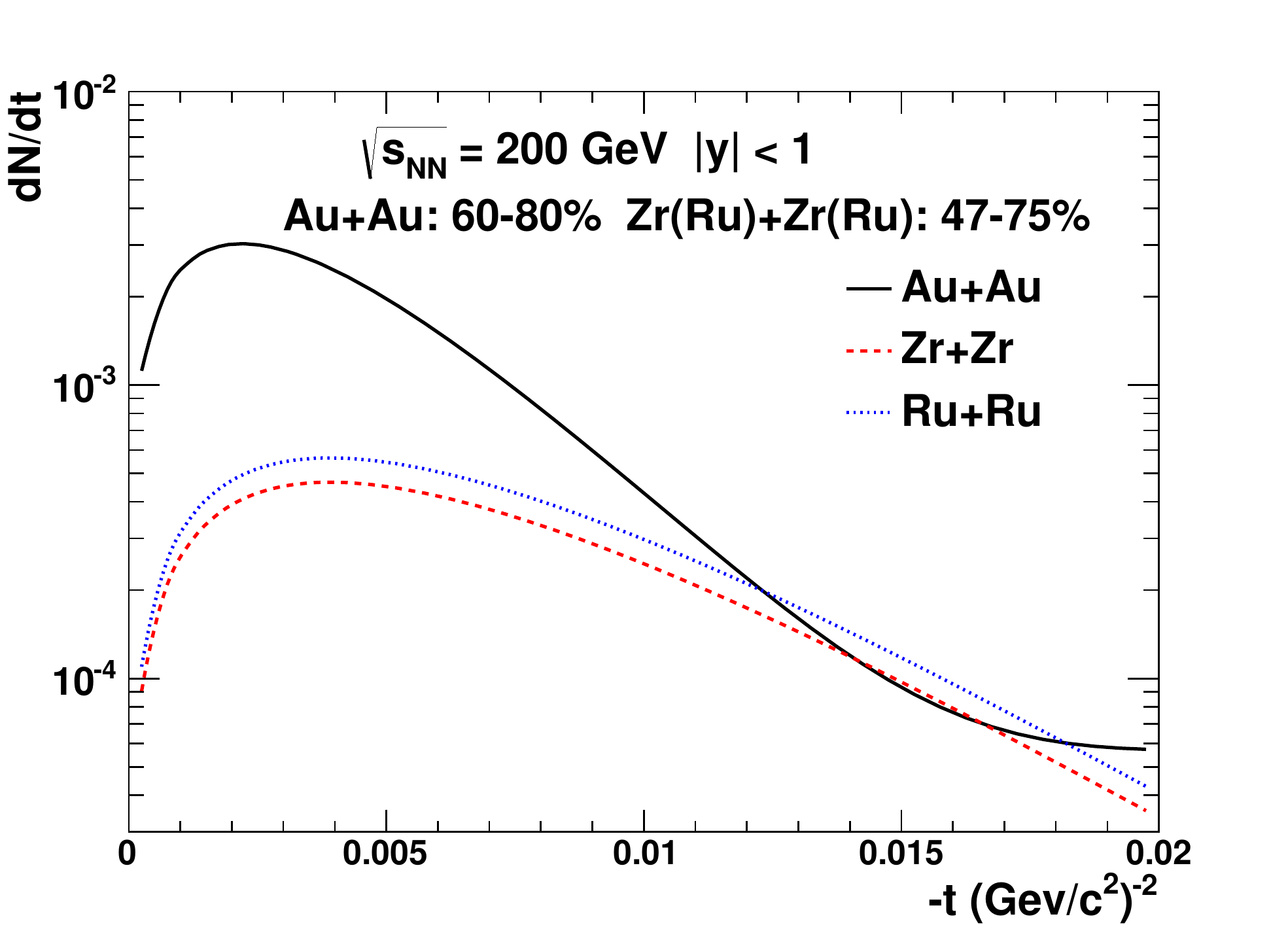}
\caption{The $t$ distribution of coherently produced J/$\psi$ at $\sqrt{s_{\rm{NN}}} =$ 200 GeV in Au+Au collisions for $60-80\%$ centrality class (solid line), Ru+Ru collisions for $47-75\%$ centrality class (dotted line), and Zr+Zr collisions for $47-75\%$ centrality class (dashed line). }
\label{figure3}
\end{figure}

  \renewcommand{\floatpagefraction}{0.75}
\begin{figure}[htbp]
\includegraphics[keepaspectratio,width=0.45\textwidth]{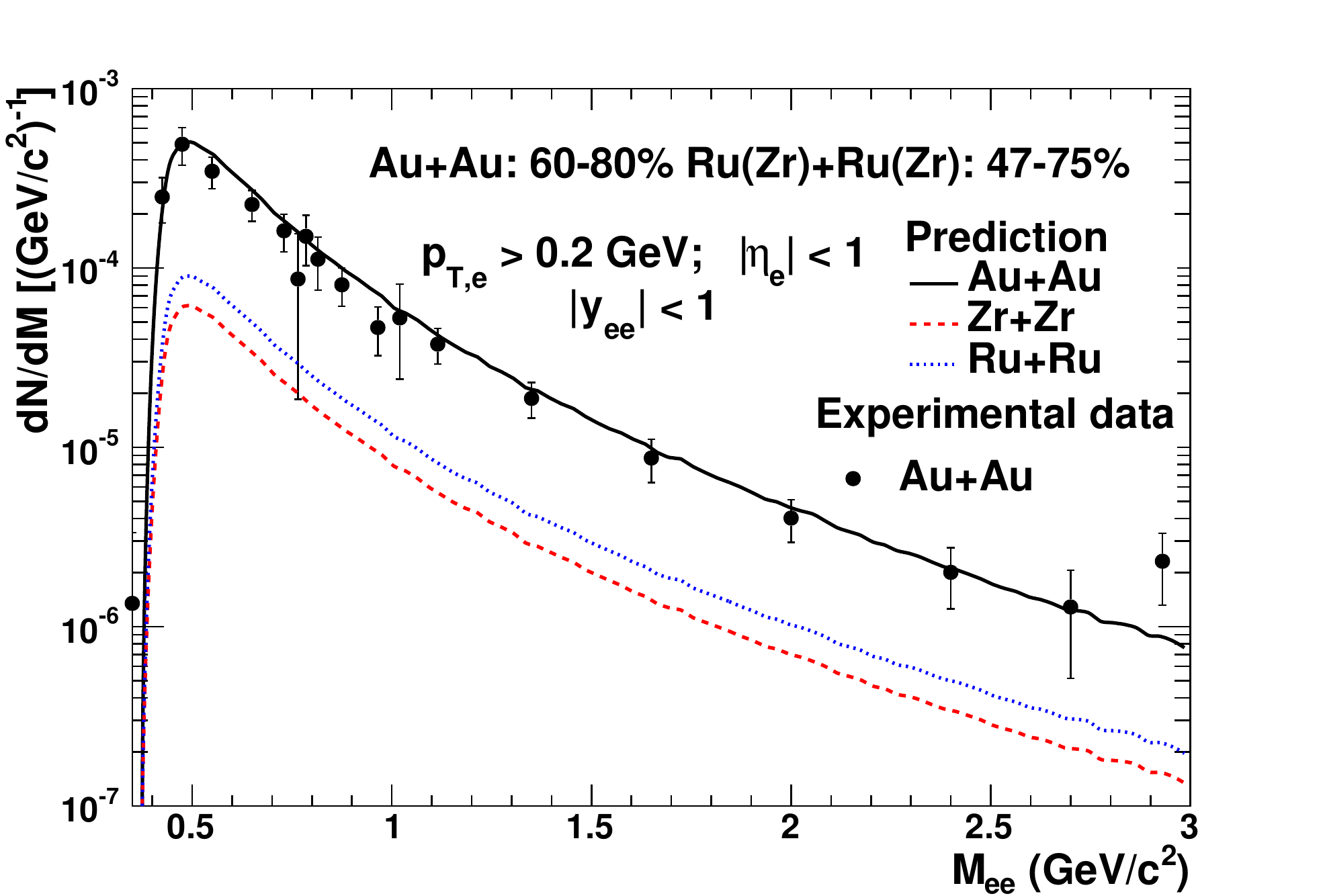}
\caption{The invariant mass spectrum of electron-positron pair at $\sqrt{s_{\rm{NN}}} =$ 200 GeV in Au+Au collisions for $60-80\%$ centrality class (solid line), Ru+Ru collisions for $47-75\%$ centrality class (dotted line), and Zr+Zr collisions for $47-75\%$ centrality class (dashed line). The experimental measurements~\cite{PhysRevLett.121.132301} in $60-80\%$ centrality class from STAR are also plotted for comparison.The results are filtered to match the fiducial acceptance described in the text.}
\label{figure4}
\end{figure}
Figure~\ref{figure2} shows the coherent J/$\psi$ yields, including interference effect, as a function of number of participants ($N_{\rm{part}}$) at $\sqrt{s_{\rm{NN}}} =$ 200 GeV in Au+Au (solid line), Ru+Ru (dotted line), and Zr+Zr (dashed line) collisions. The predictions in Au+Au and isobaric collisions all follow a trend from rise to decline towards central collisions. The increasing of cross section from peripheral to semi-peripheral collisions results from the larger photon flux induced by nucleus with smaller impact parameter. However, the later on inversion of trend originates from the destructive interference and external disturbance from overlap region, which prevails over the increasing photon flux towards central collisions. The turning point of the trend in Au+Au collisions is at higher $N_{\rm{part}}$ value than those of isobaric collisions, which is due to the collision geometry and nucleus profile differences. The production rate in Ru+Ru collisions is 1.2 times that in Zr+Zr collisions, following the exact $Z^{2}$ scaling. And the yields in isobaric collisions is dramatically smaller than that in Au+Au collisions, which is conducted by the huge $Z$ difference combined with different nuclear profile and collision geometry. The significant  differences in production rate among the three collision systems lead to dramatically different enhancements with respect to hadronic background, which provides us a sensitive probe to test the coherent photoproduction in HHIC.

The differential distributions of coherently produced J/$\psi$ in the three collision systems are also studied in this paper. Fig.~\ref{figure3} shows the $t$ distributions of coherently produced J/$\psi$ at $\sqrt{s_{\rm{NN}}} =$ 200 GeV in Au+Au collisions for $60-80\%$ centrality class (solid line), Ru+Ru collisions for $47-75\%$ centrality class (dotted line), and Zr+Zr collisions for $47-75\%$ centrality class (dashed line). The Mandelstam variable t is expressed as $t = t_{\parallel} +t_{\perp}$, with $t_{\parallel} = -M_{\text{J}/\psi}^{2}/(\gamma^{2}e^{\pm y})$ and $t_{\perp} = - p_{T}^{2}$. At top RHIC energies, $t_{\parallel}$ is very small and neglected here($t \simeq -p_{T}^{2}$). The specified centrality class is chosen to guarantee the same hadronic backgrounds in the three collisions systems. The rapid drops towards $p_{T} \rightarrow 0$ in $t$ distributions come from the destructive interference from the two colliding nuclei, while the downtrends at relative higher $p_{T}$ range reveal diffraction pattern, which are determined by the nuclear density distributions. The corresponding $t$ value at peak positron in Au+Au collisions is smaller than those in isobaric collisions, which is due to the larger impact parameter in Au+Au collision for the selected centrality class. And the slope of the down trend at relative higher $p_{T}$ range in Au+Au collisions is deeper than those in isobaric collisions, owning to larger nuclear profile for Au nucleus. There is no difference in $t$ distributions between isobaric collisions, since we use the same nuclear density distributions for Zr+Zr and Ru+Ru.

In comparison to coherent photon-nucleus interactions, the $Z^{4}$ dependence of coherent photon-photon process make it a more significant signal to be tested in HHIC. The cross-section of photon-photon interactions is heavily concentrated among near-threshold pairs, which are not visible to existing detectors. So, calculations of the total cross-section are not particularly useful. Instead, we calculate the cross-section and kinematic distributions with acceptances that match that used by STAR. Fig.~\ref{figure4} shows the invariant mass spectrum of electron-positron pair at $\sqrt{s_{\rm{NN}}} =$ 200 GeV in Au+Au collisions for $60-80\%$ centrality class (solid line), Ru+Ru collisions for $47-75\%$ centrality class (dotted line), and Zr+Zr collisions for $47-75\%$ centrality class (dashed line). The results are filtered to match the fiducial acceptance at STAR: daughter track transverse momentum $p_{T} >$ 0.2 GeV/c, track pseudo-rapidity $|\eta| <$ 1, and pair rapidity $|y| <$ 1. The experimental measurements~\cite{PhysRevLett.121.132301} in $60-80\%$ centrality class from STAR are also shown for comparison, which could be reasonaly described by our calculation. The mass distribution shapes for the three collision systems are almost the same, while the relative yield ratios are 7.9 : 1.5 : 1.0 for Au+Au, Ru+Ru, and Zr+Zr collisions, respectively. The large differences provide excellent experimental feasibility to test the production mechanism in isobaric collisions.
\section{summary}
In summary, we have performed calculations of J/$\psi$ production from coherent photon-nucleus interactions and electron-positron pair production from coherent photon-photon interactions in hadronic isobaric collisions. We show that the production rate of the coherent photon products at RHIC top energy differ significantly between isobaric collisions, and in comparison with those in Au+Au collisions, the differences grow enormous. The differential $t$ distributions of coherently produced J/$\psi$ in isobaric collisons are also studied, which possesses different shapes compared with that in Au+Au collisions due to the different nuclear profile and collison geometry. The predictions for isobaric collisions carried out in this paper provides theoretical basline for the further experimental test, which would be performed in the near future.
\section{acknowledgement}
We thank Dr. Spencer Klein and Prof. Pengfei Zhuang for useful discussions. This work was funded by the National Natural Science Foundation of China under Grant Nos. 11775213, 11505180 and 11375172, the U.S. DOE Office of Science under contract No. DE-SC0012704, and MOST under Grant No. 2014CB845400.
\nocite{*}
\bibliographystyle{aipnum4-1}
\bibliography{aps}
\end{document}